\documentclass[usenatbib]{mn2e}
\usepackage{graphicx}
\usepackage{times}
\def\gtrsim{~\rlap{$>$}{\lower 1.0ex\hbox{$\sim$}}}
\def\ltsim{~\rlap{$<$}{\lower 1.0ex\hbox{$\sim$}}}

\title[Resolved optical line emission in the Superantennae]
    {Spectroscopically- and spatially-resolved optical line 
    emission in the Superantennae (IRAS 19254-7245)\thanks{Based on 
    observations collected at the European Organisation for Astronomical 
    Research in the Southern Hemisphere, Chile [080.B-0085].}}
\author[G. J. Bendo et al.]
    {George J. Bendo$^1$, David L. Clements$^1$, Sophia A. Khan$^2$ \\
    $^1$ Astrophysics Group, Imperial College London, Blackett Laboratory, 
    London SW7 2AZ, United Kingdom\\
    $^2$ ALMA Fellow, Pontificia Universidad Cat\'{o}lica, Departamento de 
    Astronom\'{i}a y Astrof\'{i}sica, 4860 Vicu\~{n}a Mackenna Casilla 306,
    Santiago 22, Chile}
\date{}
\pagerange{\pageref{firstpage}--\pageref{lastpage}}
\pubyear{}

\begin{document}
\label{firstpage}
\maketitle

\begin{abstract}
We present VIMOS integral field spectroscopic observations of the
ultraluminous infrared galaxy (ULIRG) pair IRAS 19254-7245 (the
Superantennae).  We resolve H$\alpha$, [N {\small II}], [O {\small
I}], and [S {\small II}] emission both spatially and spectroscopically
and separate the emission into multiple velocity components.  We
identify spectral line emission characteristic of star formation
associated with both galaxies, broad spectral line emission from
the nucleus of the southern progenitor, and potential outflows with
shock-excited spectral features near both nuclei.  We estimate that
$\ltsim10$\% of the 24~$\mu$m flux density originates from star
formation, implying that most of the 24~$\mu$m emission originates
from the AGN in the southern nucleus.  We also measure a gas consumption
time of $\sim1$~Gyr, which is consistent with other measurements of
ULIRGs.
\end{abstract}


\section{Introduction}

We are undertaking an integral field spectroscopic survey of a
volume-limited sample of 18 ultraluminous infrared galaxies (ULIRGs;
galaxies where $L_{IR}>10^{12}L_\odot$) with the Visible Multi-Object
Spectrograph \citep[VIMOS;][]{letal03} at the Very Large Telescope.
We have two main scientific goals.  First, we want to study gas inflow
and outflow in ULIRGs, with particular emphasis on the detection of
starburst-driven superwinds or AGN-driven jets, and determine whether
these outflows may be part of a feedback mechanism that inhibits star
formation.  Second, we want to examine gas excitation mechanisms
within these galaxies and determine whether AGN or star formation
dominate the energetics of ULIRGs.  The results from this survey can
be applied to understanding the dynamics and rest-frame optical
spectra of both nearby and more distant ULIRGs, especially in light of
recently published integral field spectroscopic observation of
infrared- and submillimetre-luminous $z>1$ objects
\citep[e.g.][]{setal05, scslbbil06}.

We present here the first results from this survey: VIMOS integral
field spectroscopic observations of IRAS 19254-7245 (the
Superantennae).  The object, which has two distinct nuclei and two
tidal tails that extend over hundreds of kpc \citep{mlm91}, has been
the subject of many detailed studies.  The southern nucleus contains
an AGN, as shown by studies at multiple wavelengths \citep{vblmpw02,
  cetal02, betal03, bffbp03, rti07}, but the northern nucleus appears
to be completely dominated by star formation \citep{betal03, bffbp03}.
Some evidence has been given for the presence of several distinct
dynamical components within the inner arcmin of the galaxy, including
a broad line component (full-width half-maximum (FWHM)
$\sim2000$-2500~km~s$^{-1}$) associated with the southern nucleus,
narrower components (FWHM $<500$~km~s$^{-1}$) associated with the
progenitor discs, and some high-velocity clouds \citep{vblmpw02,
  rti07}.  However, the previous optical studies have been mainly
single slit spectra that have lacked spatial information, and previous
near-infrared integral field spectroscopic observations covered only
an $8\times8$~arcsec region that does not include all the emission
from the southern disc.  The VIMOS data that we present here, which
covers the central $35 \times 35$~arcsec, allows us to resolve optical
line emission both spatially and spectroscopically, so we can clearly
discern the AGN, star-forming regions, and other structures within
both galaxies.  We used this galaxy to test many of the analysis
methods that will be applied to the sample as a whole.  The spectral
line emission in this galaxy pair was detected at significantly larger
radii and the line-emitting structures are more complex than most
other sample galaxies, which is why we have focused on it for this
first paper.

\begin{figure*}
\includegraphics[width=57mm]{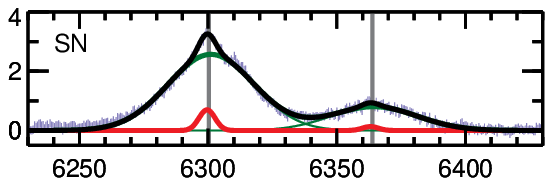}
\includegraphics[width=57mm]{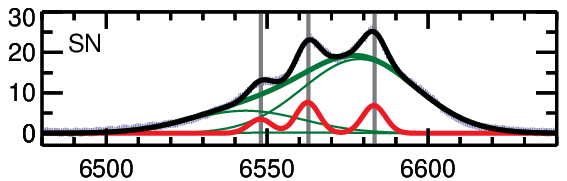}
\includegraphics[width=57mm]{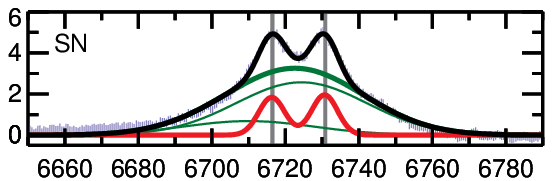}
\includegraphics[width=57mm]{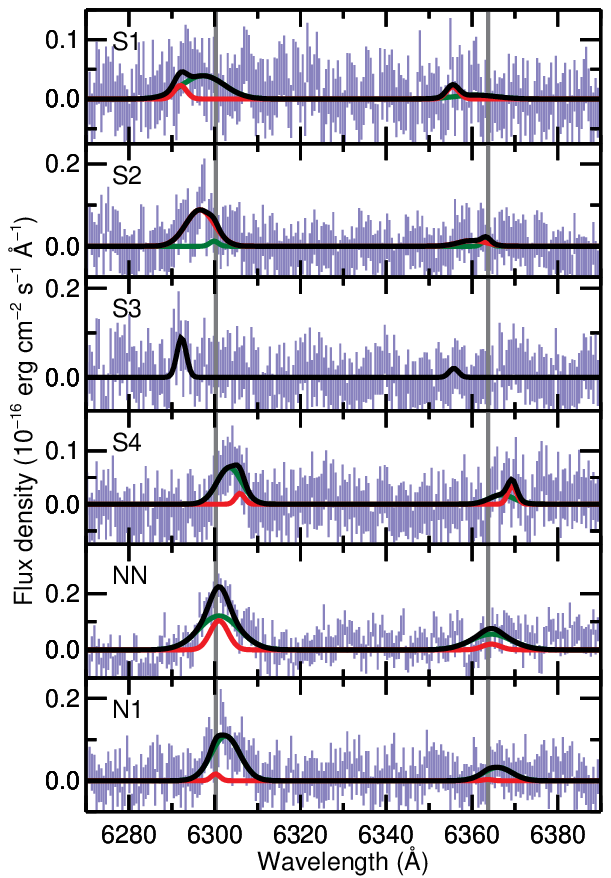}
\includegraphics[width=57mm]{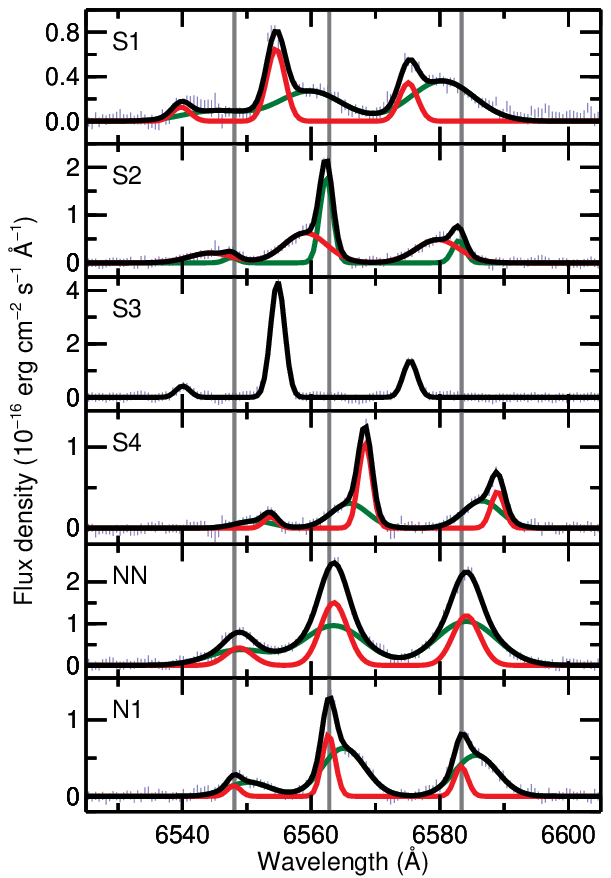}
\includegraphics[width=57mm]{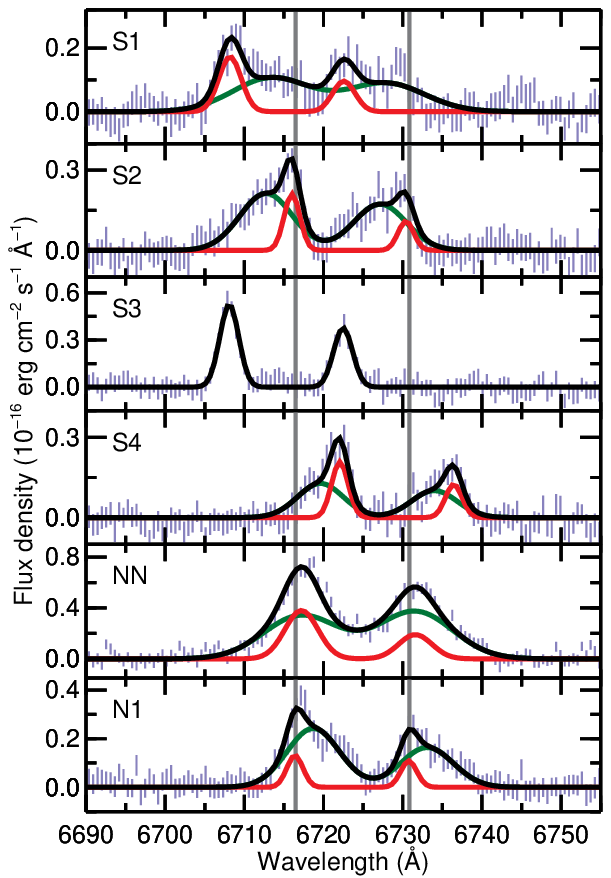}
\caption{Subsections of the spectra extracted from different regions
  drawn as blue error bars, with the best fitting models overplotted
  as solid black lines.  When two velocity components are fit to the
  data, component 1 is shown in red and component 2 is shown in green.
  For region SN, the individual spectral lines for the broader
  velocity component are shown as thin green lines.  Line centers for
  the [O {\small I}] 6300 and 6364~\r{A} lines, the [N {\small II}]
  6548 and 6583~\r{A} lines, the 6563~\r{A} H$\alpha$ line, and the
  6716 and 6731~\r{A} lines are overplotted as grey lines.  Parameters
  from these fits are given in Table~\ref{t_line}.  The spectra for
  region SN are shown over a broader wavelength range because the
  lines are broader than in other regions.}
\label{f_line}
\end{figure*}

\begin{table*}
\centering
\begin{minipage}{147mm}
\caption{Parameters for line fits for regions in Figure~\ref{f_map}
    \label{t_line}}
\small{
\begin{tabular}{@{}cccccccc@{}}
\hline
    &
    H$\alpha$ &
    H$\alpha$ &
    \multicolumn{5}{c}{Flux (erg cm$^{-2}$ s$^{-1}$)} \\
    Region &
    velocity$^a$ &
    line width$^b$ &
    [O {\small I}] &
    H$\alpha$ &
    [N {\small II}] &
    [S {\small II}] &
    [S {\small II}] \\
    &
    (km s$^{-1}$) &
    (km s$^{-1}$) &
    6300 \r{A} &
    6563 \r{A} &
    6583 \r{A} &
    6716 \r{A} &
    6731 \r{A} \\
\hline
 &
  $-7.9 \pm 0.3$ &
  $379.8 \pm 0.9$ &
  $5.4 \pm 0.3$ &
  $67.8 \pm 0.3$ &
  $60.0 \pm 0.3$ &
  $16.27 \pm 0.16$ &
  $17.32 \pm 0.17$\\
  \raisebox{1.5ex}[0pt]{SN}\vspace{0.7ex} &
  $-162.9 \pm 1.7$ &
  $1851.3 \pm 1.5$ &
  $107.3 \pm 0.25$ &
  $59 \pm 3$ &
  $761.9 \pm 1.5$ &
  $53.9 \pm 0.6$ &
  $97.9 \pm 0.6$ \\
&
  $-379 \pm 4$ &
  $153 \pm 7$ &
  $0.08 \pm 0.07$ &
  $2.3 \pm 0.2$ &
  $1.23 \pm 0.14$ &
  $0.61 \pm 0.09$ &
  $0.34 \pm 0.08$\\
  \raisebox{1.5ex}[0pt]{S1}\vspace{0.7ex} &
  $-137 \pm 14$ &
  $529 \pm 17$ &
  $0.48 \pm 0.16$ &
  $3.3 \pm 0.2$ &
  $4.5 \pm 0.3$ &
  $1.31 \pm 0.17$ &
  $1.10 \pm 0.16$\\
&
  $-21 \pm 5$ &
  $108 \pm 5$ &
  $0.03 \pm 0.05$ &
  $4.5 \pm 0.4$ &
  $1.15 \pm 0.15$ &
  $0.53 \pm 0.10$ &
  $0.27 \pm 0.08$ \\
  \raisebox{1.5ex}[0pt]{S2}\vspace{0.7ex} &
  $-170 \pm 11$ &
  $358 \pm 9$ &
  $0.74 \pm 0.15$ &
  $5.2 \pm 0.4$ &
  $4.0 \pm 0.2$ &
  $1.78 \pm 0.16$ &
  $1.44 \pm 0.15$\\
S3\vspace{0.7ex} &
  $-366.3 \pm 0.7$ &
  $117.3 \pm 0.7$ &
  $0.24 \pm 0.08$ &
  $11.85 \pm 0.14$ &
  $3.75 \pm 0.07$ &
  $1.6 \pm 0.3$ &
  $1.13 \pm 0.19$\\
&
  $256 \pm 6$ &
  $114 \pm 6$ &
  $0.05 \pm 0.06$ &
  $2.8 \pm 0.4$ &
  $1.2 \pm 0.2$ &
  $0.55 \pm 0.13$ &
  $0.33 \pm 0.09$ \\
  \raisebox{1.5ex}[0pt]{S4}\vspace{0.7ex} &
  $140 \pm 20$ &
  $311 \pm 16$ &
  $0.50 \pm 0.15$ &
  $2.2 \pm 0.5$ &
  $2.4 \pm 0.3$ &
  $0.9 \pm 0.2$ &
  $0.73 \pm 0.15$ \\
&
  $34 \pm 3$ &
  $226 \pm 13$ &
  $0.5 \pm 0.2$ &
  $7.9 \pm 1.4$ &
  $6.3 \pm 1.3$ &
  $2.0 \pm 0.5$ &
  $1.0 \pm 0.3$ \\
  \raisebox{1.5ex}[0pt]{NN}\vspace{0.7ex} &
  $33 \pm 4$ &
  $510 \pm 20$ &
  $1.4 \pm 0.4$ &
  $11 \pm 3$ &
  $12 \pm 3$ &
  $4.0 \pm 0.9$ &
  $4.4 \pm 0.7$ \\
&
  $-4 \pm 6$ &
  $110 \pm 8$ &
  $0.04 \pm 0.13$ &
  $2.1 \pm 0.4$ &
  $1.0 \pm 0.2$ &
  $0.33 \pm 0.10$ &
  $0.28 \pm 0.09$ \\
  \raisebox{1.5ex}[0pt]{N1}\vspace{0.7ex} &
  $102 \pm 10$ &
  $356 \pm 8$ &
  $0.90 \pm 0.15$ &
  $5.2 \pm 0.5$ &
  $4.5 \pm 0.3$ &
  $2.01 \pm 0.18$ &
  $1.36 \pm 0.15$ \\
\hline
\end{tabular}
$^a$ These are calculated relative to a central velocity of 
    17950 km s$^{-1}$.  Negative velocities correspond to blueshifted lines. \\
$^b$ These are based on the FWHM of the lines.
}
\end{minipage}
\end{table*}
\normalsize

\section{Observations, data reduction}

Observations were performed with the VIMOS integral field unit at the
Very Large Telescope on 2007 October 9.  The data were taken using the
HR red grism (6300-8700~\r{A}), which has a spectral resolution of
3100, and a plate scale of 0.67~arcsec for each fibre that gave a
field of view of $27 \times 27$~arcsec for each pointing.  The seeing
during the observations was $\sim0.7$~arcsec.  Spectra were measured
in five pointings offset from each other by 5~arcsec, which provided
redundancy within the central region and ensures the availability of
blank sky for background measurements.  This strategy is used for all
galaxies in the ULIRG survey.

The VIMOS data reduction pipeline produces calibrated spectra that
show the spectra measured for each fibre.  The data from each pointing
are stored in four files that contain the spectra from the four
quadrants of the integral field unit.  We used these
pipeline-processed data for our analysis, but additional processing
was needed to transform the spectra into a data cube.  First, for each
pointing and for each quadrant, we identified fibres that measured
background emission by integrating the spectra and using an iterative
process to remove fibres with high continuum signals, which would
indicate the presence of emission from the target.  Median background
spectra for each quadrant and each pointing were determined using
these background fibres, and these background spectra were subtracted
from the data.  Following this, the spectra from each pointing were
mapped into individual spectral cubes, and then the cubes were median
combined to produce the final spectral cube.

\section{Spectral line fitting and analysis}

Before measuring spectral line emission, we corrected the data for
redshift using a velocity of $\sim17950$ km s$^{-1}$, which was
estimated empirically to be the approximate velocity of narrow-line
emission from the southern nucleus, and we subtracted a continuum
determined by fitting a line through the data between 6100-6200~\r{A}
and between 6800-7000~\r{A} in the rest frame.  We then fit Gaussian
functions to the H$\alpha$, [O {\small I}], [N {\small II}], and
[S~{\small II}] lines within each spectrum in the data cube.  We
forced the fits to treat the offset between adjacent spectral features
(i.e. adjacent [O {\small I}] lines, adjacent [S {\small II}], or
lines near H$\alpha$) as constants.  We generally determined whether
to fit one or two velocity components by visually inspecting the lines
for features such as skewed line profiles or double-peaked structures
that would indicate that two velocity components are present, but we
also decided to fit two velocity components to the data when the two
component fit produced a lower reduced $\chi^2$.  When we fit one
velocity component, we used the same line widths for adjacent lines.
When we fit two velocity components to the data, we fit both
simultaneously.  To reduce the uncertainties in the line fits where
two velocity components were present, we performed fits to all
spectral features between 6200-6800~\r{A} where the offsets among all
spectral lines were treated as constants and where the corresponding
line widths for each velocity component were treated as equal.  We
also forced all line fluxes to be positive in all fits.

\begin{figure}
\begin{center}
\includegraphics[height=74.5mm]{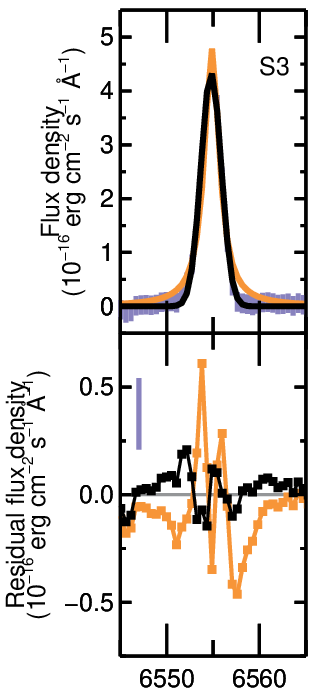}
\includegraphics[height=74.5mm]{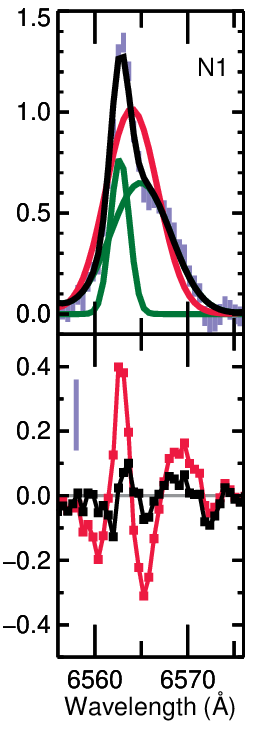}
\includegraphics[height=74.5mm]{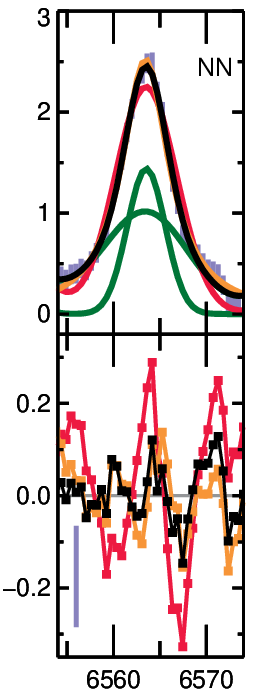}
\end{center}
\caption{The top panels show spectra for the regions around the
  H$\alpha$ line extracted from the regions in Figure~\ref{f_map}.
  The data are shown as blue error bars.  The best fitting model (one
  Gaussian function for S3 and two Gaussian functions for N1 and NN)
  are shown in black.  Individual components of the two Gaussian
  function models are shown in green.  Alternate single Lorentz
  functions for regions S3 and NN are shown in orange (although the
  orange lines are mostly overlapped by the black lines); alternative
  single Gaussian functions for regions N1 and NN are shown in
  red. The bottom panels show the residuals from the best fitting
  models in black and the residuals from the alternate models in red
  or orange, with the individual data points shown as squares and the
  typical size of the error bars shown as a blue line on the left side
  of the panels.}
\label{f_line_robust}
\end{figure}

Examples of spectral line fits are shown in Figure~\ref{f_line}, with
parameters for the best fit lines given in Table~\ref{t_line}.  Each
spectrum was extracted from multiple spatial pixels for the analysis
later in this section, but similar fits were applied to the spectra
for single spatial pixels.  We also used three of these regions to
show in Figure~\ref{f_line_robust} examples of the robustness of the
fits to the H$\alpha$ line.  Region S3 is an example or where we
observed spectral line emission from a single velocity component; we
demonstrate with this profile that the spectral line is fit better by
a Gaussian function than by a Lorentz function.  Region N1 is an
example of where the spectral lines are significantly skewed and must
be fit by two Gaussian functions.  Region NN is a special case found
in only one location in the object where the lines exhibit extra
kurtosis.  When we fit one Gaussian function, we saw that the lines
had both high central peaks and broad wings when compared to the fit,
as shown by the high positive residuals for the H$\alpha$ line fit in
Figure~\ref{f_line_robust}.  The line is better fit by either a single
Lorentz function or by two Gaussian functions with similar central
wavelengths.  However, the emission from region N1 immediately to the
west of NN consists of skewed lines with two velocity components that
have line widths and line ratios similar to those in the two Gaussian
component fit to NN.  This suggest that the line profiles for NN
should be modeled as two Gaussian components and not one Lorentz
component.

\begin{figure*}
\includegraphics[width=40mm]{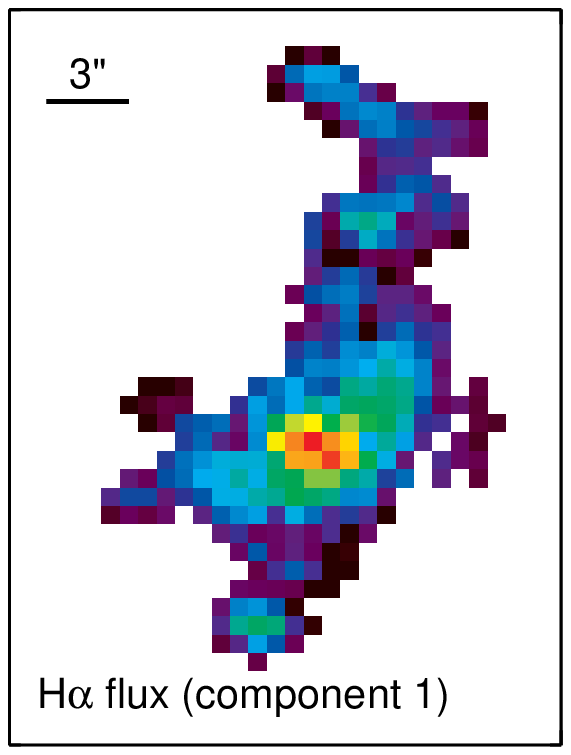}
\includegraphics[width=40mm]{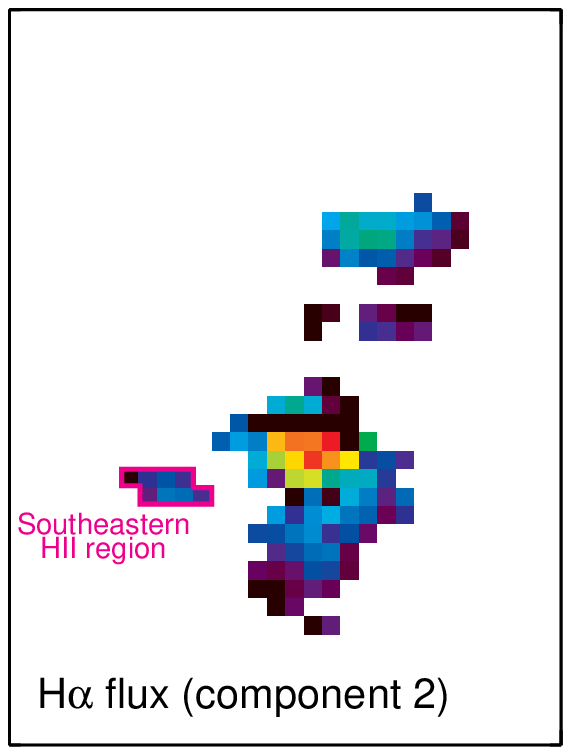}
\includegraphics[width=40mm]{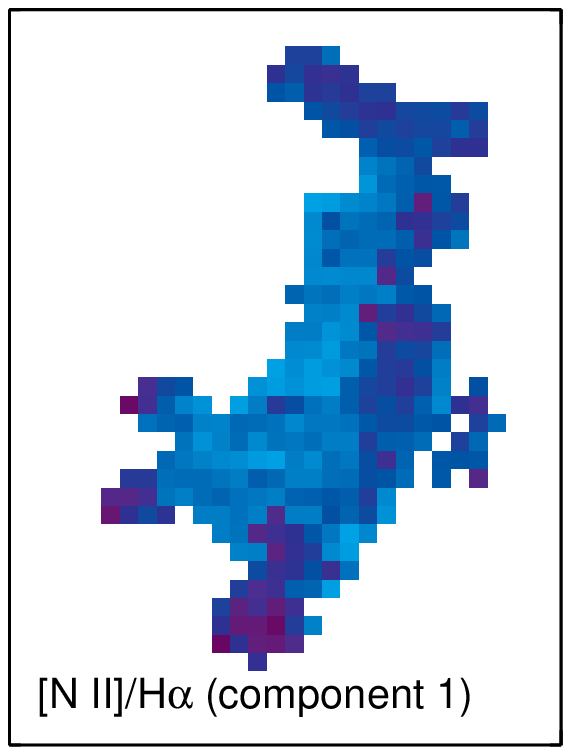}
\includegraphics[width=40mm]{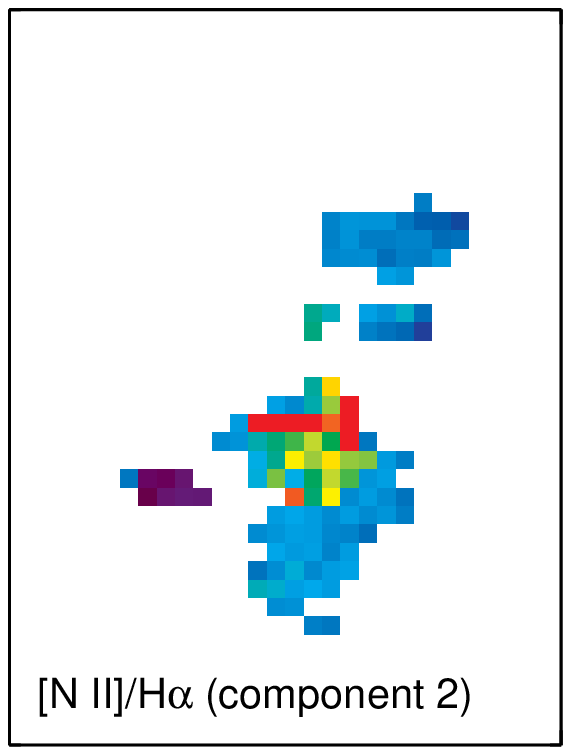}\\
\includegraphics[width=80.8mm]{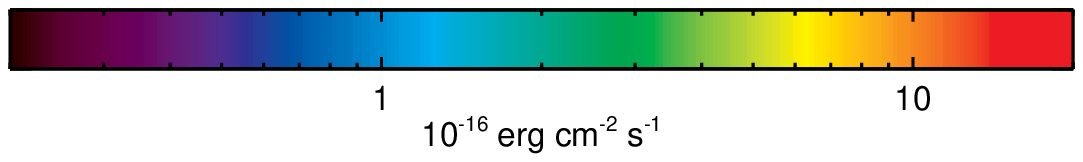}
\includegraphics[width=80.8mm]{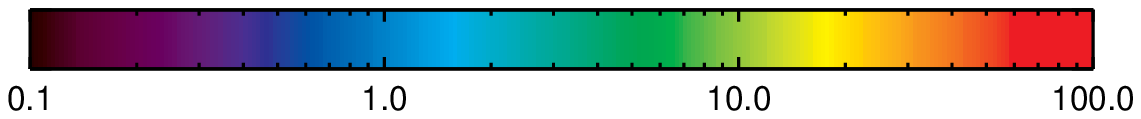}\\
\includegraphics[width=40mm]{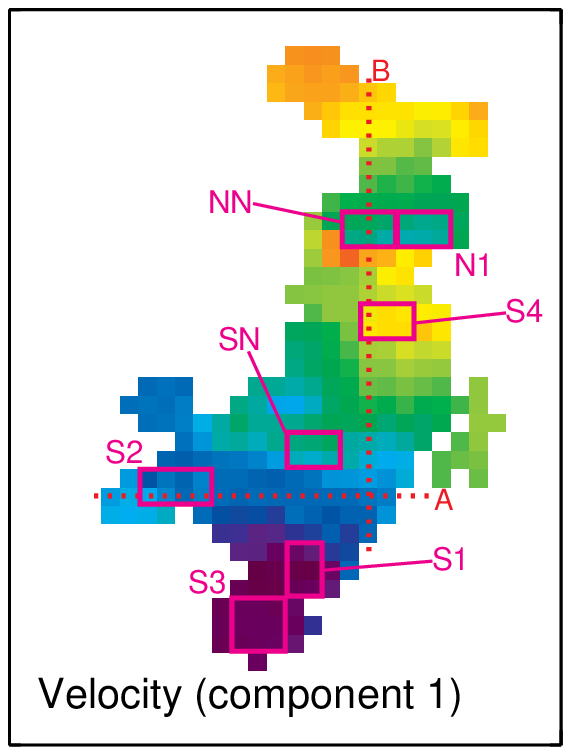}
\includegraphics[width=40mm]{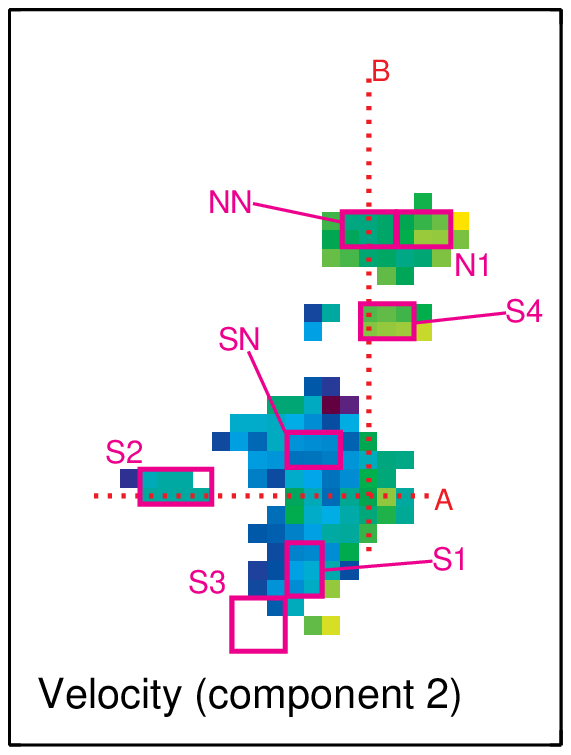}
\includegraphics[width=40mm]{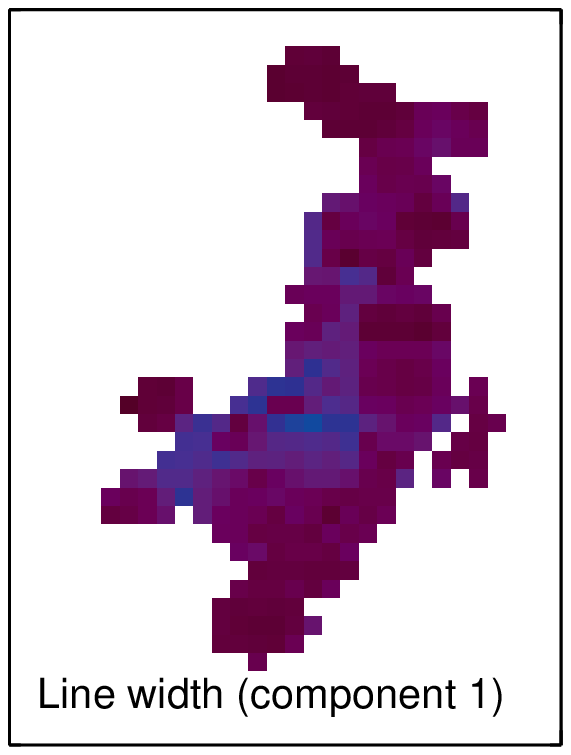}
\includegraphics[width=40mm]{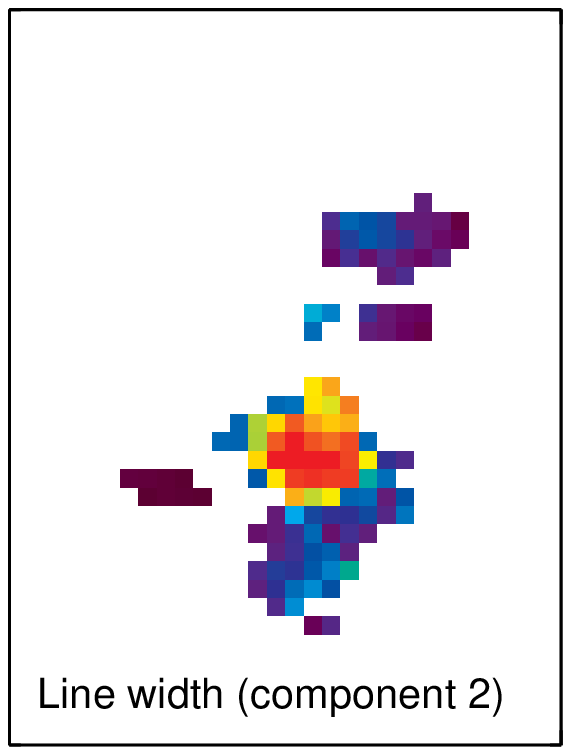}\\
\includegraphics[width=80.8mm]{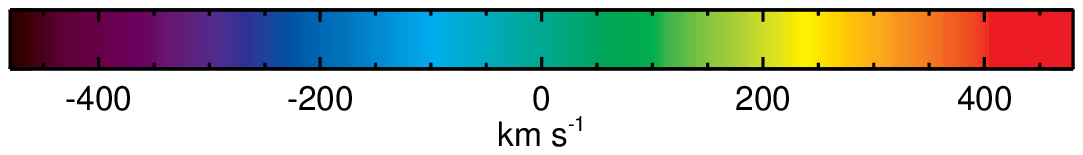}
\includegraphics[width=80.8mm]{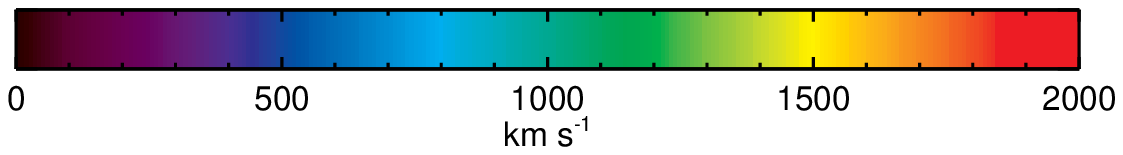}
\caption{Images of the H$\alpha$ flux, the [N {\small II}]/H$\alpha$
  flux ratio, the H$\alpha$ velocity, and the H$\alpha$ line widths
  for the inner $20 \times 26$~arcsec of the Superantennae.  The
  kinematic structures have been divided into two components, with the
  broader spectral lines generally placed in the second component; see
  the text for details.  The objects in each frame may not necessarily
  be physically associated.  North is up and east is left in each
  image, and the angular scale is shown in the H$\alpha$ flux image
  for the first component.  The magenta regions in the velocity images
  are the locations for which spectra are displayed in
  Figure~\ref{f_line}.  The red dotted lines show the locations with
  velocities and line widths plotted in Figure~\ref{f_axes}. }
\label{f_map}
\end{figure*}

\begin{figure}
\begin{center}
\includegraphics{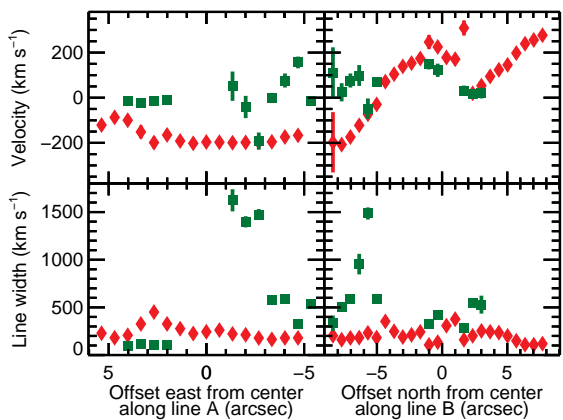}
\end{center}
\caption{Velocities and line widths for component 1 (as red diamonds)
  and component 2 (as green squares) for the locations shown by the
  dotted lines in Figure~\ref{f_map}.}
\label{f_axes}
\end{figure}

The results from the fits to each spatial pixel were then used to
produce maps of the parameters, as shown in Figure~\ref{f_map}.
Additionally, Figure~\ref{f_axes} shows how the velocities and line
widths vary across vertical and horizontal lines that were placed so
as to cover locations of interest with two velocity components.
Component 1 contains almost all of the emission from locations with
only a single line component and most of the narrower line emission in
locations with two line components.  Component 2 generally contains
broader spectral line emission.  In the southeastern H{\small II}
region, however, we found that the broader spectral line component had
velocities that were closer to the velocities traced by the narrow
line component in adjacent pixels and that the velocities of the
narrow component were sharply different from component 1 in adjacent
pixels, as seen in Figure~\ref{f_axes}.  We therefore shifted the
narrower line emission for these locations into component 2 and the
broader line emission into component 1.  We also found six pixels
along the southern or western edge of the detected region where either
a single line component or a narrower line component had velocities
closer to that for the broader line component in adjacent pixels, and
so the narrower or single line component was shifted to component 2 in
these cases as well.

We can see two rotating progenitors and a couple of H$\alpha$-bright
knots in component 1 in Figure~\ref{f_map}.  In component 2, we can
see broad (FWHM $\sim1850$~km~s$^{-1}$) spectral line emission from
the nucleus, a broad arc to the southeast of the nucleus, and a few
other regions in both progenitors. The velocity maps and
Figure~\ref{f_axes} clearly show that the two progenitors are rotating
in the same direction.

The individual spectra shown in Figure~\ref{f_line} are useful for
understanding the nature of the different regions of the galaxy.  In
the following discussion, we use the spectral line diagnostics
presented by \citet{kgkh06} to determine whether the spectral line
emission is consistent with star formation or AGN-like emission,
although the usefulness of these diagnostics is limited without
additional measurements of H$\beta$ and [O {\small III}] (5007~\r{A})
fluxes.  We treat locations where log([O{\small
    I}]~(6300~\r{A})/H$\alpha$)~$<$~-1.0, log([N{\small
    II}]~(6583~\r{A})/H$\alpha$)~$<$~0 and log([S{\small
    II}]~(6716,31~\r{A})/H$\alpha$)~$<$~-0.2 as being dominated by
photoionisation from star formation, whereas regions with higher line
ratios are treated as having AGN-like emission.  The southern nucleus
(SN) contains both broader and narrower emission line components.  The
narrow component probably originates from the progenitor disc remnant,
and the spectral line ratios are consistent with star formation.  The
broader component is consistent with AGN activity.  In particular, the
H$\alpha$ emission is very weak compared to other spectral features,
most notably seen in the map of the [N {\small II}]/H$\alpha$ ratio,
indicating that the gas is ionised by a hard radiation field.  This is
consistent with results for Pa$\alpha$ found by \citet{rti07}.  S1
contains both a narrower component associated with the disc remnant
and a broader component that is part of the arc extending from the
southern nucleus.  The line ratios of the broader component are
consistent with AGN-like emission, which implies that this extension
is a shock similar to those seen in other ULIRGs by \citet{mac06}.
The apparent physical connection to the southern nucleus, the smooth
changes in velocity from the southern nucleus to the arc, and the
uniform [N {\small II}]/H$\alpha$ ratio across the arc imply that it
is gas ejected from the southern nucleus.  However, the orientation
and motion of the arc relative to the rotation of the southern disc
remnant imply that it could be a tidal feature.  The broader component
in S2 is associated with the southern disc remnant, while the narrower
component is associated with a cloud moving at a higher velocity.
Both components in region S2 have line ratios that are consistent with
photoionisation.  The spectrum of the region in S3 is consistent with
star formation.  The spectrum of S4 contains two components, and the
broader component is consistent with AGN emission.  In NN and N1, the
line ratios of the narrower components are like H{\small II} emission,
but the broader components are AGN-like.  While the broader components
of NN and N1 may be associated with ejecta from the southern nucleus
or with material stripped or ejected from the southern disc remnant,
its location and velocity imply that the emission originates from gas
ejected from the northern disc remnant.

Based on the application of the diagnostics from \citet{kgkh06},
component 1 and the southeastern H{\small II} region in component 2
(outlined in magenta in Figure~\ref{f_map}) appear to trace all of the
star formation.  The total H$\alpha$ flux from star formation as
traced by these components is $(3.52\pm 0.02)\times10^{-14}$ erg
s$^{-1}$ cm$^{-2}$, or $\sim70$\% of the total H$\alpha$ flux from the
galaxy.  To understand the relative contribution of star formation to
the 24~$\mu$m flux density, we can use extinction measurements from
the literature and the H$\alpha$ flux in
\begin{equation}
f_{H\alpha}e^{0.812A_V}=f_{H\alpha}+0.022\frac{c}{24\mu m}f_{24\mu m}
\end{equation}
(adapted from \citet{zetal08}) to estimate the 24~$\mu$m flux density
from star formation.  In this equation, $f_{H\alpha}$ is the measured
H$\alpha$ flux, $f_{H\alpha}e^{0.812A_V}$ represents the
extinction-corrected H$\alpha$ flux, and $f_{24\mu m}$ is the
24~$\mu$m flux density.  Although similar equations have been
published for individual H{\small II} regions within galaxies, the
\citet{zetal08} version is the only one currently published that has
been calibrated for use on global flux measurements, and given that
extreme starbursts and AGN with $L_{IR}$ up to $10^{12}$~L$_\odot$
were included in their analysis, their relation should be applicable
to this galaxy.  The extinction function of \citet{sm79} with $R=3.1$
was used to derive the extinction correction term on the left side of
the equation.  The $A_V$ measured in the southern nucleus by
\citet{vblmpw02} and \citet{bffbp03} using single-slit spectra is
$\sim 3.1$.  Assuming that this extinction applies to all star-forming
regions and that the extinction is intrinsic to the source itself, we
estimate the 24~$\mu$m flux density from star-forming regions to be
$\sim0.15$~Jy.  In contrast, the total 25~$\mu$m flux density measured
by IRAS is $1.24\pm0.06$~Jy \citep{metal89}.  We tentatively conclude
that $\gtrsim90$\% of the 24~$\mu$m flux originates from something
other than star formation; the AGN in the southern nucleus is the most
likely source of this emission.  

Our results are consistent with the
spectral energy distribution (SED) modelling by \citet{bffbp03}, which
predicts that the AGN is the dominant source of 25~$\mu$m emission in
the Superantennae.  While \citet{faerfc03} found using SED template
fitting that AGN may be the primary source of 25~$\mu$m emission in
ULIRGs in general, we disagree with their conclusions that the
Superantennae SED can be explained entirely by star formation.  The
starburst template that they used does not accurately describe the
Superantennae data between 10-100~$\mu$m, so it may be an inaccurate
description of the 24~$\mu$m flux density.  While the 24~$\mu$m flux
density may be dominated by an AGN, star formation may represent
approximately half of 1-1000~$\mu$m flux, and that starburst emission
may dominate the far-infrared emission from the Superantennae, as
predicted by \citet{bffbp03}.  Our results rely on the assumption that
the kinematic components that we have identified are the only
locations with star formation and that the extinction across the
star-forming regions in this galaxy is not variable and accurately
represented by $A_V\simeq3.1$.  \citet{bffbp03} found that extinction
was lower in the outer regions of the galaxy, which could reduce the
estimated 24~$\mu$m flux density from star formation.

Using the extinction-corrected H$\alpha$ flux for star-forming
regions, a distance of 245~Mpc (calculated using a velocity of 17950
km s$^{-1}$ and $H_0$ of 73 km s$^{-1}$ Mpc$^{-1}$), and the
conversion of H$\alpha$ flux to star formation rate given by
\citet{k98}, we estimate the star formation rate to be
25~M$_\odot$~yr$^{-1}$.  The total molecular gas mass has been
measured to be $1.9-3.0\times10^{10}$~M$_\odot$ \citep{mbgjs90,
  vblmpw02}, which would imply a gas consumption time of $\sim1$~Gyr.
While this gas consumption time does not account for gas recycling, it
is still indicative of the efficiency of star formation.  This is
slightly high but still consistent with the ``few times $10^8$~yr''
gas consumption times estimated for ULIRGs using star formation rates
derived from far-infrared data \citep{tetal06}, but, as expected, it
is lower than the $\sim3$~Gyr gas consumption times measured for
normal spiral galaxies \citep{ktc94}.

\section*{Acknowledgements}

We thank the referees for their useful comments on this paper.  This
work was funded by STFC.  SAK thanks FONDECYT for support through
Proyecto No. 1070992.

\label{lastpage}

\end{document}